\documentclass[aps,prl,twocolumn,showpacs,nofootinbib,floatfix]{revtex4-1}

\pdfoutput=1
\usepackage{amssymb}
\usepackage{graphicx}
\usepackage{amsmath}
\usepackage{wasysym}
\usepackage{verbatim}

\begin{document}
\title{Comment on ``Joint Anisotropy and Source Count Constraints on the Contribution of Blazars to the Diffuse Gamma-Ray Background''} 

\author{J.\ Patrick Harding$^{1,2}$} \email{hard0923@umd.edu}
\author{Kevork N.\ Abazajian$^1$} \email{kevork@uci.edu}
\affiliation{$^1$Department of Physics \& Astronomy, University of
  California, Irvine, California, 92697} \affiliation{$^2$Maryland
  Center for Fundamental Physics, Department of Physics, University of
  Maryland, College Park, Maryland 20742}

\begin{abstract}
We show the conclusions claimed in the manuscript arXiv:1202.5309v1 by
Cuoco, Komatsu and Siegal-Gaskins (CKS) are not generally valid.  The
results in CKS are based on a number of simplifying assumptions
regarding the source population below the detection threshold and the
threshold flux itself, and do not apply to many physical models of the
blazar population.  Physical blazar population models that match the
measured source counts above the observational threshold can account
for $\sim$60\% of the diffuse gamma-ray background intensity between 
1-10 GeV, while the assumptions in CKS limit the intensity to 
$\lesssim$30\%.  The shortcomings of the model considered in CKS arise 
from an over-simplified blazar source model.  A number of the simplifying
assumptions are unjustified, including: first, the adoption of an
assumed power-law source-count distribution, $dN/dS$, to arbitrary low
source fluxes, which is not exhibited in physical models of the blazar
population; and, second, the lack of blazar spectral information in
calculating the anisotropy of unresolved gamma-ray blazar emission.
We also show that the calculation of the unresolved blazars'
anisotropy is very sensitive to the spectral distribution of the
unresolved blazars through the adopted source resolution threshold
value, and must be taken into account in an accurate anisotropy
calculation.
\end{abstract}

\maketitle

\section{\vskip -0.6cm Introduction}
The contribution of unresolved blazars to the diffuse gamma-ray
background (DGRB) has been of interest for some time (see
Ref.~\cite{Abazajian:2010pc}, hereafter ABH, for a discussion). The
recent manuscript by Cuoco, Komatsu, and
Siegal-Gaskins~\cite{Cuoco:2012yf} (hereafter CKS) has derived limits
on the contribution of blazars to the DGRB from a combination of
measurements of the DGRB anisotropy, source-count distribution, and
intensity. Using a simplistic $dN/dS$ for the blazars, and neglecting
any blazar spectral information, the CKS analysis concludes that
blazars can contribute no more than $30\%$ of the DGRB intensity,
independent of the measured angular correlation power in the DGRB. In
this note, we show that the CKS limit on the blazar contribution to
the DGRB intensity is not generally valid, and strictly the result of
their chosen over-simplified model.  Such a model neglects many
crucial features of physically-motivated blazar models,
e.g. Refs.~\cite{Abazajian:2010pc,Cavadini:2011ig,Singal:2011yi,Stecker:2010di,Ando:2006cr,Inoue09}. 
Importantly, using a physically-constrained source-count distribution
{\it above the threshold} that is consistent with that measured by the
Fermi-LAT collaboration~\cite{Collaboration:2010gqa} (hereafter FB10)
and assumed by CKS, ABH find an intensity contribution to the DGRB
between 1-10 GeV of approximately 60\%, in direct contradiction to the
general claim in CKS of a required $\lesssim 30\%$ contribution.  In
this note, we summarize the reasons for this discrepancy, which reside
in a number of invalid assumptions in CKS.

\section{Blazar Flux Source-Count Distribution Function}

\begin{figure}[t]
\begin{center}
\includegraphics[width=3.4truein]{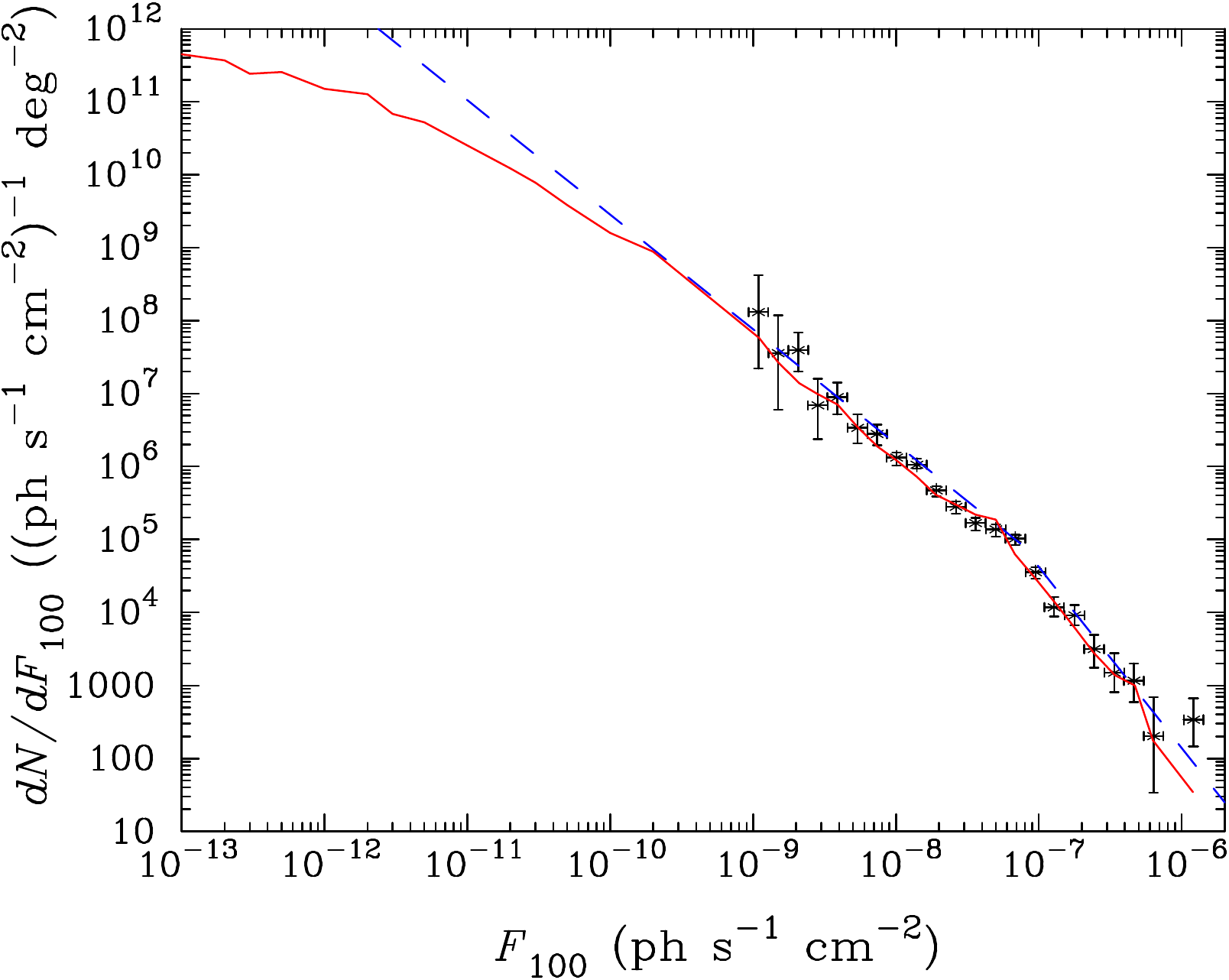}
\end{center}
\caption{The best-fit model for the source-count distribution function
  $dN/dF_{100}$ for the ABH blazar model (solid red line). The data are
  from FB10. The power-law fit to 
  $dN/dF_{100}$ from FB10 is shown for comparison (dashed 
  blue line). Note that in this plot, 
  $F_{100}=\int_{100\rm\ MeV}^{\infty}(dN/dE)\ dE$ for comparison 
  to the data, instead of $S$ as defined in the text.
\label{blazardNdS}}
\end{figure}

The blazar source count distribution functions $dN/dS$ give the number
of blazars expected at a particular flux $S$, defined in this note as
\begin{equation}
S=\int_{1\rm\ GeV}^{10\rm\ GeV}dE\ \frac{dN}{dE}\enspace,
\end{equation}
which is consistent with the flux $S$ defined in CKS, though $S$ can
be defined in different energy bands. 
The Large Area Telescope (LAT)
on the Fermi Gamma-Ray Space Telescope has measured the $dN/dS$ for
blazars to be consistent with a broken power-law over the Fermi-LAT
sensitivity range in FB10.  CKS assumes that the full blazar $dN/dS$
follows this faint-end single power-law, down to zero flux. They
calculate the diffuse blazar intensity in the $1-10$ GeV band as
\begin{equation}
I = \int_0^{S_{\rm t}}{\frac{dN}{dS}\ S \ dS}\label{I_eqn}\enspace,
\end{equation}
and they determine the value of $S_{\rm t}$, the ``flux sensitivity
threshold'' below which all sources are undetected by the Fermi-LAT,
by normalizing this intensity integral to the measured blazar
intensity from FB10. This definition of $S_{\rm t}$ neglects the
strong spectral dependence of the Fermi-LAT point source sensitivity.
CKS then calculates the Poisson term of the angular power spectrum for
undetected blazars as
\begin{equation}
C_P = \int_0^{S_{\rm t}}{\frac{dN}{dS}\ S^2 \ dS}\label{C_P_eqn}\enspace,
\end{equation}
and compares this value to the measured $C_P$ in the DATA:CLEANED
sample from Table II of Ref.~\cite{Ackermann:2012uf}. Using the limits
on $I$ and $C_P$, they conclude that ``unresolved blazars account for
only 30\% of the IGRB intensity but 100\% of the angular
power.''\footnote{CKS refers to the DGRB as the isotropic diffuse
  gamma-ray background (IGRB).} However, the CKS analysis makes
several simplifications which change these results greatly when
examining physical blazar models.  Importantly, CKS incorrectly use
the same value of $S_t$ in calculating $C_P$, equation~\eqref{C_P_eqn}, as
used in $I$, equation~\eqref{I_eqn}, even though the point source exclusion
limit for the former is the 1FGL catalog, $TS=25$, and that for the
latter DGRB intensity is more conservative, $TS=50$.  

Unlike the broken power-law $dN/dS$ used by CKS, at low fluxes $dN/dS$
is expected to flatten rather than continuing to increase down to zero
flux. The blazar model of ABH, for example, exhibits a flattening of
source-count distribution at low fluxes, as shown in
figure~\ref{blazardNdS}. The ABH model was determined by using a
luminosity-dependent density evolution (LDDE) model of the gamma-ray
luminosity function using the spectral energy distribution (SED)
sequence for blazars, which was fit to the Fermi-LAT source-count
distribution of FB10 and the Fermi-LAT-measured DGRB of
Ref.~\cite{Abdo:2010nz} using the spectrally-dependent flux limit of
sources. Rather than continuing to rise as a power-law for low fluxes,
this source-count distribution flattens and provides less blazars at
low flux than a simple power-law extrapolation would predict, as is
shown in figure~\ref{blazardNdS}. Additionally, a power-law
extrapolation down to zero flux of the type assumed in CKS is
mathematically inconsistent, giving a divergent number of blazars
within our cosmological horizon, $N = \int_0^{S_{\rm t}} (dN/dS) dS$,
while the physical source-count distribution of ABH does not.

Aside from a change in the number of sources at low flux, the
definition of the flux itself leads to a $dN/dS$ at low fluxes which
is different from the broken power-law used by CKS. Consider the
intensity coming from near the threshold flux $S_{\rm t}$: $S_1 <
S_{\rm t} < S_2$
\begin{equation}
I_{\rm band} = \int_{S_1}^{S_2}{\frac{dN}{dS}\ S \ dS}\enspace.
\end{equation}
Let us consider the sources' fluxes to be from a population that changes
its spectrum and/or number density from a distribution $N_1$ to $N_2$
near the threshold flux,
\begin{eqnarray}
S = 
\begin{cases}
\int_{E_1}^{E_2}{(dN_1/dE)\ dE} \qquad \text{for $S\le S_{\rm t}$}
\cr\ \\
\int_{E_1}^{E_2}{(dN_2/dE)\ dE} \qquad \text{for $S > S_{\rm t}$}
\end{cases}
\enspace.
\end{eqnarray}
The flux from the sources is spectrally dependent, and could be, e.g.,
two different average power law spectral distributions
\begin{equation}
\frac{dN_i}{dE} = N_0 \left(\frac{E}{E_0}\right)^{-\Gamma_i}\enspace.
\end{equation}
The intensity calculation therefore changes its form from just below
to just above the threshold, and the former lacks any dependence on
the latter,
\begin{eqnarray}
I_{\rm band} &=& \int_{S_1}^{S_{\rm t}}{\frac{dN_1}{dS}\ S\ dS}
 + \int_{S_{\rm t}}^{S_2}{\frac{dN_2}{dS}\ S \ dS}\enspace.
\end{eqnarray}
Therefore the contribution from below the threshold has, in general,
complete independence from that above the threshold, and it should be
clear that the contribution from just {\it below} the threshold can be
very different from just {\it above} the threshold. Therefore, {\it
  all predictive power of equation~\ref{I_eqn} (and therefore in
  CKS) is lost.} The same problem can be illustrated with
equation~\ref{C_P_eqn}, which was used by CKS to calculate $C_P$.

Equation~\ref{I_eqn} is only valid when blindly extrapolating a
power-law for the source population to fluxes below the observable
flux threshold, and furthermore requires the assumption of an
invariant source population spectrum below that threshold.  Though
neither of these assumptions is valid in general blazar source
population models, CKS has adopted both.  What is properly needed in
understanding the nature of the source population below the threshold
flux is a physical picture of the blazar population.  One such
physical model is provided by the LDDE plus SED sequence model in ABH.
\section{Diffuse Intensity, Angular Correlation, and the Threshold Flux}
In addition to the simplistic assumptions used in extrapolating $dN/dS$
below the Fermi-LAT sensitivity threshold, CKS mishandles the
calculation of the threshold itself. In CKS, the threshold flux
$S_{\rm t}$ is calculated using equation~\ref{I_eqn} and normalizing
$I$ to the measured intensity reported in FB10. The calculation of
$C_P$ is then made using the $S_{\rm t}$ calculated from $I$. However,
$C_P$ is highly sensitive to barely-unresolved sources near the
threshold, so the calculation of $C_P$ is strongly dependent on the
chosen value of $S_{\rm t}$. A factor of two change in $S_{\rm t}$
only changes $I$ by $\sim20\%$ but can change $C_P$ by a factor of three.

CKS additionally considers the blazar model of
Ref.~\cite{Stecker:2010di} and calculates $C_P$ for this
model. However, they use the $S_{\rm t}$ previously calculated for the
FB10 blazar model, which was normalized to a significantly different
value of $I$ than Ref.~\cite{Stecker:2010di} calculates. This flux
threshold value is not the correct one for the
Ref.~\cite{Stecker:2010di} blazar model, and therefore, the
CKS-calculated value of $C_P$ for this model is not valid. For
comparison, for the ABH blazar model, we find that the blazar
intensity from 1-10 GeV is
$2.2\times10^{-7}\rm\ ph\ cm^{-2}\ s^{-1}\ sr^{-1}$, approximately
$60\%$ of the DGRB. For this intensity, we find a flux threshold of
$S_{\rm t}=2.9\times10^{-9} \rm\ ph\ cm^{-2}\ s^{-1}$, which is much
different than the $S_{\rm t}=3.7\times10^{-10}
\rm\ ph\ cm^{-2}\ s^{-1}$ from CKS.

Additional problems with CKS are related to equations~\ref{I_eqn}
and~\ref{C_P_eqn} for the intensity and anisotropy of the source
population. There is an inherent integration and averaging over the
source spectrum in these expressions.  Figure 1 of FB10 shows the
threshold flux for the Fermi-LAT to be not a single flux value but
rather a strong function of the blazar spectral index. Depending on
the blazar index, the threshold flux can vary by two orders of
magnitude. This is important because the blazar intensity $I$ is more
sensitive to hard sources than the blazar anisotropy $C_P$, so the
threshold flux $S_{\rm t}$ is, in general, not the same for the
calculation of $I$ and the calculation of $C_P$.

As a simple example, the blazar model of FB10, which extrapolates the
Fermi-LAT broken power-law $dN/dS$ below the Fermi-LAT threshold,
considers a spread in the blazar spectral indices $\Gamma$:
\begin{eqnarray}
\frac{dN}{dSd\Gamma}&=& f(S)\ g(\Gamma)\\
f(S)&=&
  \left\{ 
    \begin{array}{cl} 
      A\ S^{-\beta_1} &\ S\ge S_b \\
      A\ S_b^{-\beta_1+\beta_2}\ S^{-\beta_2} &\ S<S_b
    \end{array} 
    \right.\\
g(\Gamma)&=&\exp\left[-\frac{(\Gamma-\mu)^2}{2\sigma^2}\right]\enspace.
\end{eqnarray}
Including the blazar index distribution, equations~\ref{I_eqn} and~\ref{C_P_eqn} become
\begin{eqnarray}
I&=&\int_{-\infty}^{\infty}d\Gamma\int_0^{S_{\rm t}(\Gamma)}dS\ S\ {\frac{dN}{dSd\Gamma}}\\
C_P&=&\int_{-\infty}^{\infty}d\Gamma\int_0^{S_{\rm t}(\Gamma)}dS\ S^2\ {\frac{dN}{dSd\Gamma}}
\end{eqnarray}
Using the $0.1-100$ GeV band model from table 4 of FB10 and the
threshold fluxes in FB10 figure 1, we calculate $I$ and $C_P$ for this
blazar model. Note that this analysis was done using
$F_{100}=\int_{100\rm\ MeV}^{\infty}(dN/dE)\ dE$ rather than
$S$, to be consistent with FB10 figure 1. The FB10 model gives
$I=2.4\times10^{-6}\rm\ ph\ cm^{-2}\ s^{-1}\ sr^{-1}$ and
$C_P=3.7\times10^{-14}\rm\ (ph\ cm^{-2}\ s^{-1})^2\ sr^{-1}$. Using
equations~\ref{I_eqn} and~\ref{C_P_eqn}, the equivalent
index-independent threshold fluxes for each calculation are $S_{\rm
  t}(I)=1.7\times10^{-8}\ {\rm ph\ cm^{-2}\ s^{-1}}$ and $S_{\rm
  t}(C_P)=3.8\times10^{-8}\ {\rm ph\ cm^{-2}\ s^{-1}}$. As shown
above, the large difference in $S_{\rm t}$ significantly affects the
calculation of $C_P$. CKS fails to  take this effect into account.

For a full LDDE plus SED sequence blazar model, like ABH, the
flux-dependence of the blazar spectrum must also be taken into
account. An LDDE-based blazar model which integrates over blazar
luminosity and redshift has been considered, but only with a simple
power-law energy spectrum, rather than the full blazar
SED~\cite{Ando:2006cr}. To do an accurate calculation of the LDDE plus
SED model from ABH, an extension of the calculations of
Ref.~\cite{Ando:2006cr} to include the blazar SED must be
done~\cite{AnisotropyPaper}.

\section{Conclusions}
As shown in ABH and above, using a source-count distribution that is
consistent with that measured by the Fermi-LAT collaboration (FB10)
and assumed by CKS above the threshold, ABH find an intensity
contribution to the DGRB between 1-10 GeV of 60\%, in direct
contradiction to the general claim in CKS of a required $\lesssim
30\%$ contribution.  The CKS calculation of the Poisson term of the
angular power spectrum for undetected blazars is inadequate.  The
broken power-law $dN/dS$ they choose cannot be accurately extrapolated
below the Fermi-LAT flux threshold, and doing so leads to unphysical
results. CKS also use the incorrect value for the threshold flux when
calculating $C_P$ and comparing model intensity results.  Furthermore,
the model they consider fails to account for blazars' spectral
properties, which can affect the anisotropy calculation
significantly. They assume a spectrally-independent threshold flux for
the Fermi-LAT, which does not match the actual Fermi-LAT measurements.
For the other model CKS considers, from Ref.~\cite{Stecker:2010di},
they use a value of the Fermi-LAT flux threshold which does not
accurately reflect the threshold flux for that model, and therefore
this model's exclusion by the $C_P$ is questionable.

Forthcoming work should accurately consider the consistency between
angular correlations in the DGRB and its intensity, as contributed by
blazars in physically-motivated blazar models, and should not rely on
unjustified extrapolations and the other unqualified assumptions
present in CKS, as described above.

\begin{acknowledgments}
  {\it Acknowledgments---}JPH \& KNA are supported by NSF CAREER Grant
  PHY \#09-55415.
\end{acknowledgments}

\bibliography{bibliography}

\begin{thebibliography}{10}%
\makeatletter
\providecommand \@ifxundefined [1]{%
 \ifx #1\undefined \expandafter \@firstoftwo
 \else \expandafter \@secondoftwo
\fi
}%
\providecommand \@ifnum [1]{%
 \ifnum #1\expandafter \@firstoftwo
 \else \expandafter \@secondoftwo
\fi
}%
\providecommand \enquote [1]{``#1''}%
\providecommand \bibnamefont  [1]{#1}%
\providecommand \bibfnamefont [1]{#1}%
\providecommand \citenamefont [1]{#1}%
\providecommand\href[0]{\@sanitize\@href}%
\providecommand\@href[1]{\endgroup\@@startlink{#1}\endgroup\@@href}%
\providecommand\@@href[1]{#1\@@endlink}%
\providecommand \@sanitize [0]{\begingroup\catcode`\&12\catcode`\#12\relax}%
\@ifxundefined \pdfoutput {\@firstoftwo}{%
 \@ifnum{\z@=\pdfoutput}{\@firstoftwo}{\@secondoftwo}%
}{%
 \providecommand\@@startlink[1]{\leavevmode\special{html:<a href="#1">}}%
 \providecommand\@@endlink[0]{\special{html:</a>}}%
}{%
 \providecommand\@@startlink[1]{%
  \leavevmode
  \pdfstartlink
   attr{/Border[0 0 1 ]/H/I/C[0 1 1]}%
   user{/Subtype/Link/A<</Type/Action/S/URI/URI(#1)>>}%
  \relax
 }%
 \providecommand\@@endlink[0]{\pdfendlink}%
}%
\providecommand \url  [0]{\begingroup\@sanitize \@url }%
\providecommand \@url [1]{\endgroup\@href {#1}{\urlprefix}}%
\providecommand \urlprefix [0]{URL }%
\providecommand \Eprint[0]{\href }%
\@ifxundefined \urlstyle {%
  \providecommand \doi [1]{doi:\discretionary{}{}{}#1}%
}{%
  \providecommand \doi [0]{doi:\discretionary{}{}{}\begingroup
  \urlstyle{rm}\Url }%
}%
\providecommand \doibase [0]{http://dx.doi.org/}%
\providecommand \Doi[1]{\href{\doibase#1}}%
\providecommand \bibAnnote [3]{%
  \BibitemShut{#1}%
  \begin{quotation}\noindent
    \textsc{Key:}\ #2\\\textsc{Annotation:}\ #3%
  \end{quotation}%
}%
\providecommand \bibAnnoteFile [2]{%
  \IfFileExists{#2}{\bibAnnote {#1} {#2} {\input{#2}}}{}%
}%
\providecommand \typeout [0]{\immediate \write \m@ne }%
\providecommand \selectlanguage [0]{\@gobble}%
\providecommand \bibinfo [0]{\@secondoftwo}%
\providecommand \bibfield [0]{\@secondoftwo}%
\providecommand \translation [1]{[#1]}%
\providecommand \BibitemOpen[0]{}%
\providecommand \bibitemStop [0]{}%
\providecommand \bibitemNoStop [0]{.\EOS\space}%
\providecommand \EOS [0]{\spacefactor3000\relax}%
\providecommand \BibitemShut [1]{\csname bibitem#1\endcsname}%
\bibitem{Abazajian:2010pc}%
  \BibitemOpen
  \bibfield{author}{%
  \bibinfo {author} {\bibfnamefont{K.~N.}\ \bibnamefont{Abazajian}}, \bibinfo
  {author} {\bibfnamefont{S.}~\bibnamefont{Blanchet}},\ and\ \bibinfo {author}
  {\bibfnamefont{J.}~\bibnamefont{Harding}},\ }%
  \bibfield{journal}{%
  \Doi{10.1103/PhysRevD.84.103007}{\bibinfo {journal} {Phys.Rev.}}\ }%
  \textbf{\bibinfo {volume} {D84}},\ \bibinfo {pages} {103007} (\bibinfo {year}
  {2011}),\ \Eprint{http://arxiv.org/abs/1012.1247}{arXiv:1012.1247
  [astro-ph.CO]}%
  \bibAnnoteFile{NoStop}{Abazajian:2010pc}%
\bibitem{Cuoco:2012yf}%
  \BibitemOpen
  \bibfield{author}{%
  \bibinfo {author} {\bibfnamefont{A.}~\bibnamefont{Cuoco}}, \bibinfo {author}
  {\bibfnamefont{E.}~\bibnamefont{Komatsu}},\ and\ \bibinfo {author}
  {\bibfnamefont{J.}~\bibnamefont{Siegal-Gaskins}}}%
   (\bibinfo {year} {2012}),\
  \Eprint{http://arxiv.org/abs/1202.5309}{arXiv:1202.5309 [astro-ph.CO]}%
  \bibAnnoteFile{NoStop}{Cuoco:2012yf}%
\bibitem{Cavadini:2011ig}%
  \BibitemOpen
  \bibfield{author}{%
  \bibinfo {author} {\bibfnamefont{M.}~\bibnamefont{Cavadini}}, \bibinfo
  {author} {\bibfnamefont{R.}~\bibnamefont{Salvaterra}},\ and\ \bibinfo
  {author} {\bibfnamefont{F.}~\bibnamefont{Haardt}}}%
   (\bibinfo {year} {2011}),\
  \Eprint{http://arxiv.org/abs/1105.4613}{arXiv:1105.4613 [astro-ph.CO]}%
  \bibAnnoteFile{NoStop}{Cavadini:2011ig}%
\bibitem{Singal:2011yi}%
  \BibitemOpen
  \bibfield{author}{%
  \bibinfo {author} {\bibfnamefont{J.}~\bibnamefont{Singal}}, \bibinfo {author}
  {\bibfnamefont{V.}~\bibnamefont{Petrosian}},\ and\ \bibinfo {author}
  {\bibfnamefont{M.}~\bibnamefont{Ajello}},\ }%
  \bibfield{journal}{%
  \bibinfo {journal} {Astrophysical Journal}}%
   (\bibinfo {year} {2011}),\
  \Eprint{http://arxiv.org/abs/1106.3111}{arXiv:1106.3111 [astro-ph.CO]}%
  \bibAnnoteFile{NoStop}{Singal:2011yi}%
\bibitem{Stecker:2010di}%
  \BibitemOpen
  \bibfield{author}{%
  \bibinfo {author} {\bibfnamefont{F.~W.}\ \bibnamefont{Stecker}}\ and\
  \bibinfo {author} {\bibfnamefont{T.~M.}\ \bibnamefont{Venters}},\ }%
  \bibfield{journal}{%
  \Doi{10.1088/0004-637X/736/1/40}{\bibinfo {journal} {Astrophys.J.}}\ }%
  \textbf{\bibinfo {volume} {736}},\ \bibinfo {pages} {40} (\bibinfo {year}
  {2011}),\ \Eprint{http://arxiv.org/abs/1012.3678}{arXiv:1012.3678
  [astro-ph.HE]}%
  \bibAnnoteFile{NoStop}{Stecker:2010di}%
\bibitem{Ando:2006cr}%
  \BibitemOpen
  \bibfield{author}{%
  \bibinfo {author} {\bibfnamefont{S.}~\bibnamefont{Ando}}, \bibinfo {author}
  {\bibfnamefont{E.}~\bibnamefont{Komatsu}}, \bibinfo {author}
  {\bibfnamefont{T.}~\bibnamefont{Narumoto}},\ and\ \bibinfo {author}
  {\bibfnamefont{T.}~\bibnamefont{Totani}},\ }%
  \bibfield{journal}{%
  \Doi{10.1103/PhysRevD.75.063519}{\bibinfo {journal} {Phys. Rev.}}\ }%
  \textbf{\bibinfo {volume} {D75}},\ \bibinfo {pages} {063519} (\bibinfo {year}
  {2007}),\
  \Eprint{http://arxiv.org/abs/astro-ph/0612467}{arXiv:astro-ph/0612467}%
  \bibAnnoteFile{NoStop}{Ando:2006cr}%
\bibitem{Inoue09}%
  \BibitemOpen
  \bibfield{author}{%
  \bibinfo {author} {\bibfnamefont{Y.}~\bibnamefont{Inoue}}\ and\ \bibinfo
  {author} {\bibfnamefont{T.}~\bibnamefont{Totani}},\ }%
  \bibfield{journal}{%
  \Doi{10.1088/0004-637X/702/1/523}{\bibinfo {journal} {Astrophys. J.}}\ }%
  \textbf{\bibinfo {volume} {702}},\ \bibinfo {pages} {523} (\bibinfo {year}
  {2009}),\ \Eprint{http://arxiv.org/abs/0810.3580}{arXiv:0810.3580
  [astro-ph]}%
  \bibAnnoteFile{NoStop}{Inoue09}%
\bibitem{Collaboration:2010gqa}%
  \BibitemOpen
  \bibfield{author}{%
  \bibinfo {author} {\bibfnamefont{A.~A.}\ \bibnamefont{Abdo}} \emph{et~al.}
  (\bibinfo {collaboration} {Fermi-LAT Collaboration}),\ }%
  \bibfield{journal}{%
  \Doi{10.1088/0004-637X/720/1/435}{\bibinfo {journal} {Astrophys. J.}}\ }%
  \textbf{\bibinfo {volume} {720}},\ \bibinfo {pages} {435} (\bibinfo {year}
  {2010}),\ \Eprint{http://arxiv.org/abs/1003.0895}{arXiv:1003.0895
  [astro-ph.CO]}%
  \bibAnnoteFile{NoStop}{Collaboration:2010gqa}%
\bibitem{Ackermann:2012uf}%
  \BibitemOpen
  \bibfield{author}{%
  \bibinfo {author} {\bibfnamefont{M.}~\bibnamefont{Ackermann}}, \bibinfo
  {author} {\bibfnamefont{M.}~\bibnamefont{Ajello}}, \bibinfo {author}
  {\bibfnamefont{A.}~\bibnamefont{Albert}}, \bibinfo {author}
  {\bibfnamefont{L.}~\bibnamefont{Baldini}}, \bibinfo {author}
  {\bibfnamefont{J.}~\bibnamefont{Ballet}}, \emph{et~al.}}%
   (\bibinfo {year} {2012}),\
  \Eprint{http://arxiv.org/abs/1202.2856}{arXiv:1202.2856 [astro-ph.HE]}%
  \bibAnnoteFile{NoStop}{Ackermann:2012uf}%
\bibitem{Abdo:2010nz}%
  \BibitemOpen
  \bibfield{author}{%
  \bibinfo {author} {\bibfnamefont{A.~A.}\ \bibnamefont{Abdo}} \emph{et~al.}
  (\bibinfo {collaboration} {Fermi-LAT Collaboration}),\ }%
  \bibfield{journal}{%
  \Doi{10.1103/PhysRevLett.104.101101}{\bibinfo {journal} {Phys. Rev. Lett.}}\
  }%
  \textbf{\bibinfo {volume} {104}},\ \bibinfo {pages} {101101} (\bibinfo {year}
  {2010}),\ \Eprint{http://arxiv.org/abs/1002.3603}{arXiv:1002.3603
  [astro-ph.HE]}%
  \bibAnnoteFile{NoStop}{Abdo:2010nz}%
\bibitem{AnisotropyPaper}%
  \BibitemOpen
  \bibfield{author}{%
  \bibinfo {author} {\bibfnamefont{K.~N.}\ \bibnamefont{Abazajian}}\ and\
  \bibinfo {author} {\bibfnamefont{J.}~\bibnamefont{Harding}},\ }%
  \bibinfo {howpublished} {in preparation} (\bibinfo {year} {2012})%
  \bibAnnoteFile{NoStop}{AnisotropyPaper}%
\end{thebibliography}%
\end{document}